\documentclass[sigconf,nonacm]{acmart}

\usepackage{enumitem}
\usepackage{threeparttable}
\usepackage{multirow}
\usepackage{makecell}
\usepackage{pgfplots}
\usepackage{subcaption}
\usepackage{balance}

\AtBeginDocument{%
  }

\begin{document}

\title{A Macro- and Micro-Hierarchical Transfer Learning Framework for Cross-Domain Fake News Detection}

% \author{Anonymous Author(s)}

\author{Xuankai Yang}
\affiliation{%
  \department{School of Computing}
  \institution{Macquarie University}
  \city{Sydney}
  \country{Australia}}
\orcid{0000-0002-4985-3560}
\email{xuankai.yang@students.mq.edu.au}

\author{Yan Wang}
\authornote{Corresponding author.}
\affiliation{%
  \department{School of Computing}
  \institution{Macquarie University}
  \city{Sydney}
  \country{Australia}}
\orcid{0000-0002-5344-1884}
\email{yan.wang@mq.edu.au}

\author{Xiuzhen Zhang}
\affiliation{%
  \department{School of Computing Technologies}
  \institution{RMIT University}
  \city{Melbourne}
  \country{Australia}}
\orcid{0000-0001-5558-3790}
\email{xiuzhen.zhang@rmit.edu.au}

\author{Shoujin Wang}
\affiliation{%
  \department{Data Science Institute}
  \institution{University of Technology Sydney}
  \city{Sydney}
  \country{Australia}}
\orcid{0000-0003-1133-9379}
\email{shoujin.wang@uts.edu.au}

\author{Huaxiong Wang}
\affiliation{%
  \department{School of Physical and Mathematical Sciences}
  \institution{Nanyang Technological University}
  \city{Singapore}
  \country{Singapore}}
\orcid{0000-0002-7669-8922}
\email{hxwang@ntu.edu.sg}

\author{Kwok Yan Lam}
\affiliation{%
  \department{College of Computing and Data Science}
  \institution{Nanyang Technological University}
  \city{Singapore}
  \country{Singapore}}
\orcid{0000-0001-7479-7970}
\email{kwokyan.lam@ntu.edu.sg}

\renewcommand{\shortauthors}{Xuankai Yang et al.}

\begin{abstract}
  Cross-domain fake news detection aims to mitigate domain shift and improve detection performance by transferring knowledge across domains. Existing approaches transfer knowledge based on news content and user engagements from a source domain to a target domain. However, these approaches face two main limitations, hindering effective knowledge transfer and optimal fake news detection performance. Firstly, from a micro perspective, they neglect the negative impact of veracity-irrelevant features in news content when transferring domain-shared features across domains. Secondly, from a macro perspective, existing approaches ignore the relationship between user engagement and news content, which reveals shared behaviors of common users across domains and can facilitate more effective knowledge transfer. To address these limitations, we propose a novel macro- and micro- hierarchical transfer learning framework (MMHT) for cross-domain fake news detection. Firstly, we propose a micro-hierarchical disentangling module to disentangle veracity-relevant and veracity-irrelevant features from news content in the source domain for improving fake news detection performance in the target domain. Secondly, we propose a macro-hierarchical transfer learning module to generate engagement features based on common users' shared behaviors in different domains for improving effectiveness of knowledge transfer. Extensive experiments on real-world datasets demonstrate that our framework significantly outperforms the state-of-the-art baselines.

\end{abstract}





% \received{20 February 2025}
% \received[revised]{12 March 2025}
% \received[accepted]{5 June 2025}

\maketitle

\section{Introduction}
In recent years, the rapid and wide spread of fake news on social media has raised increasing threats to people's daily lives~\cite{grinberg2019fake, olan2024fake, rocha2021impact}. In order to effectively combat fake news, automatically detecting fake news becomes crucial~\cite{zhou2020survey}.

\begin{figure*}[t]
  \centering
  \includegraphics[scale=0.3]{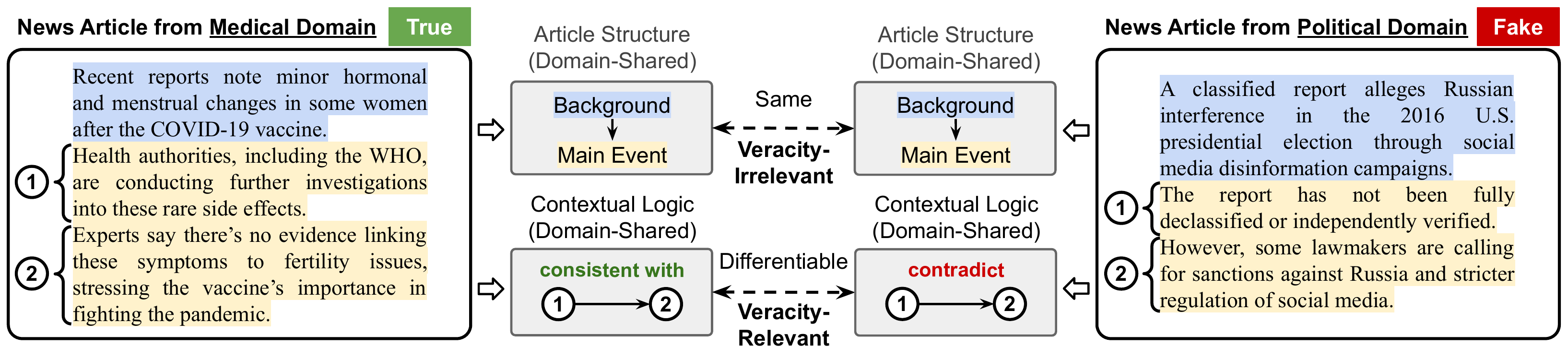}
  \caption{Examples of a news article from medical domain that is true and a news article from political domain that is fake.} 
  \label{fig:detail_examples}
  \Description{example of the micro perspective}
\end{figure*}

\begin{figure}
  \centering
  \includegraphics[scale=0.27]{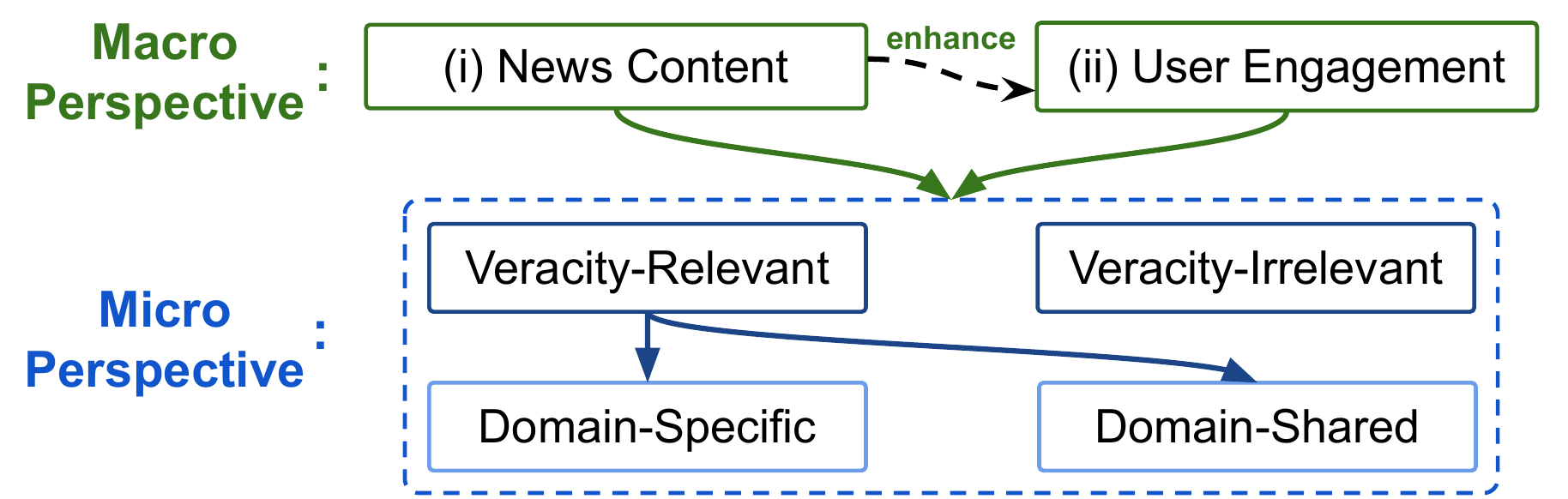}
  \caption{From the macro perspective, our approach extends the disentangling mechanism in (i) news content to enhance features accuracy in (ii) user engagement.
  %leverages both (i) news content and (ii) user engagements for knowledge transfer. 
  From the micro perspective, our approach disentangles veracity-relevant and veracity-irrelevant features before extracting domain-specific and domain-shared features for cross-domain fake news detection.}
  \label{fig:idea_examples}
  \Description{example of our novelty}
\end{figure}

In the literature, two types of data are widely utilized in fake news detection: news content~\cite{perez2017automatic, volkova2017separating, ma2016detecting, shu2019defend} and user engagements (a.k.a., user interactions)~\cite{bian2020rumor, sun2022rumor, yin2024gamc}. Based on these types of data, single-domain fake news detection has been proposed, which extracts effective features from news articles, user engagements, or both of them to identify the veracity of news (i.e., whether the news is true or fake) within a single news domain (e.g., political news or entertainment news). To mitigate the domain shift problem~\cite{pan2010survey}, cross-domain fake news detection extracts features from news content~\cite{wang2018eann, castelo2019topic, silva2021embracing, zhu2022memory, yue2023metaadapt} and user engagements~\cite{mosallanezhad2022domain, yang2024update}, and transfers them from a source domain to a target domain, thereby supporting more accurate and robust fake news detection in the target domain.

In order to achieve effective cross-domain knowledge transfer, existing cross-domain approaches are typically developed from \textbf{macro} and \textbf{micro} perspectives. From the \textit{macro perspective}, two types of data, news content and user engagement, can be utilized for detecting fake news. From the \textit{micro perspective}, to solve the issue of "what to transfer"~\cite{zang2022survey} in cross-domain knowledge transfer, news content and user engagement features can be categorized into domain-shared features (i.e., features that are common across different domains) and domain-specific features (i.e., features that are unique in a single domain). In addition, for detecting fake news, they can also be categorized into veracity-relevant features (i.e., features that differentiate between fake and true news) and veracity-irrelevant features (i.e., features that are similar in fake and true news). However, existing cross-domain approaches still face two main limitations.

\textbf{Limitation 1 (micro perspective)}: \textit{they primarily focus on extracting all domain-shared features from news content, neglecting the negative impact of veracity-irrelevant features which may also be shared across domains and thus transferred, leading to limited detection performance}. For instance, as depicted in Fig.~\ref{fig:detail_examples}, there are two news articles from different domains following the same article structure (i.e., introduces the background before presenting the main event), thus, the article structure feature can be considered as a domain-shared feature. However, this common article structure is insufficient to distinguish true and fake news, because fake news often mimics the article structure of true news~\cite{tandoc21what}. Therefore, such domain-shared but veracity-irrelevant features do not assist in detecting fake news. In contrast, the contextual logic feature differs between true and fake news. As shown in Fig.~\ref{fig:detail_examples}, in the true news, the earlier sentence is consistent with the conclusion drawn later, while fake news often exhibits contradictions between earlier sentence and the conclusion. Therefore, such domain-shared and veracity-relevant features can help improve the detection performance.

\textbf{Limitation 2 (macro perspective)}: \textit{existing cross-domain approaches ignore the relationship between user engagement and news content, limiting their effectiveness in knowledge transfer.}
In the real world, users (also called \textit{common users} in this work) often engage with news from multiple domains~\cite{vosoughi2018spread}, exhibiting \textit{shared behaviors} across domains. Mining engagement features of these behaviors can help bridge the gap between source and target domains. Specifically, to further enhance fake news detection in the target domain, user features derived from the news they engage with in the source domain can provide additional context and enrich engagement features~\cite{chen2015opinion}. However, most of the existing cross-domain approaches only consider either news content~\cite{castelo2019topic,zhu2022memory,ran2023unsupervised, yue2022contrastive,wang2018eann, silva2021embracing,yue2023metaadapt} or user engagement~\cite{yang2024update} for knowledge transfer but ignore the relationship between them. Although Mosallanezhad et al.~\cite{mosallanezhad2022domain} considered both news content and user engagement, they simply combine them without exploring engagement features based on common users across domains.

Based on the above analyses from both macro and micro perspectives, two challenges in cross-domain fake news detection can be summarized as follows. \textbf{CH1.} \textit{From the micro perspective, how to effectively disentangle veracity-relevant and veracity-irrelevant features before extracting domain-shared and domain-specific features from news content for enabling beneficial and accurate knowledge transfer in cross-domain fake news detection?} and \textbf{CH2.} \textit{From the macro perspective, how to effectively leverage shared behaviors of common users across domains to facilitate more effective knowledge transfer?}

\textbf{Our Approach and Contributions.} To address the above two challenges, in this paper, we propose a novel \underline{M}acro- and \underline{M}icro- \underline{H}ierarchical \underline{T}ransfer learning framework (MMHT) for cross-domain fake news detection. Based on the idea depicted in Fig.~\ref{fig:idea_examples}, our framework extracts accurate and enhanced representations from news content and user engagement for fake news detection. Specifically, our framework includes two core modules: (1) \textit{micro-hierarchical disentangling module}, and (2) \textit{macro-hierarchical transfer learning module}. In particular, to address \textbf{CH1}, the first module disentangles veracity-relevant and veracity-irrelevant features from news content before extracting domain-shared and domain-specific features. In addition, to address \textbf{CH2}, the second module first derives user features from users' engagements with news and the content of engaged news, and then generate engagement features from the derived features of common users, facilitating more effective knowledge transfer.

The characteristics and contributions of this work are summarized as follows:
\begin{itemize}[leftmargin=*]
  \item We propose a novel micro-hierarchical disentangling module to address the negative impact of veracity-irrelevant features in news content for cross-domain fake news detection.
  \item We propose a novel macro-hierarchical transfer learning module that utilizes common users' shared behaviors and their engaged news in different domains to generate engagement features, thereby facilitating more effective cross-domain knowledge transfer.
  \item We conduct extensive experiments on two real-world datasets to compare our framework with eleven baselines. The results demonstrate that our MMHT significantly outperforms the best-performing baselines by an average of 4.09\% across all metrics.
\end{itemize}

\section{Related Work}
\subsection{Single-Domain Fake News Detection}
Single-domain approaches for news content-based fake news detection were initially studied. They typically extract informative features from the content of news articles to predict their veracity. For instance, studies in~\cite{perez2017automatic, volkova2017separating} analyze word usage within news articles, while studies in~\cite{shu2019defend, ma2016detecting} extract sentence-level and contextual features to detect fake news, leveraging methods such as support vector machine (SVM)~\cite{perez2017automatic}, long short-term memory (LSTM) network and convolutional neural network (CNN)~\cite{volkova2017separating}, recurrent neural network (RNN)~\cite{ma2016detecting}, and co-attention mechanism~\cite{shu2019defend}. In addition, with the development of language models, language models (LMs) such as BERT~\cite{devlin2018bert} and GPT-2, as well as large language models (LLMs) like GPT-3.5 and GPT-4, have been applied to fake news detection. Consequently, some approaches, such as studies in~\cite{hu2024bad}, have combined the strengths of both LMs and LLMs for fake news detection. 

Single-domain approaches for user engagement-based fake news detection extract additional information from user engagement to enhance the effectiveness of fake news detection. They explore features in user engagements within a single domain. For example, Nguyen et al.~\cite{nguyen2020fang} proposed a model to better learn graph representations from social context information such as user engagement for fake news detection. Khoo et al.~\cite{khoo2020interpretable} proposed a self-attention based model to extract more informative representation from both news content and user engagement. Although various approaches have been proposed, these single-domain approaches often exhibit poor fake news detection performance when applied in a news domain that is new for the approach, highlighting the issue of domain shift.

\subsection{Cross-Domain Fake News Detection}
To solve the issue of domain shift, cross-domain fake news detection approaches have been developed. Some of them focus on using different techniques to generate domain-shared features based on news content, such as machine learning methods~\cite{castelo2019topic}, neural networks~\cite{zhu2022memory}, contrastive learning~\cite{ran2023unsupervised, yue2022contrastive}, adversarial learning~\cite{wang2018eann, silva2021embracing}, as well as meta learning~\cite{yue2023metaadapt}. Cross-domain user engagement-based fake news detection approaches have also been proposed. For instance, Mosallanezhad et al.~\cite{mosallanezhad2022domain} proposed a reinforcement learning-based model to extract domain-shared features from the combination of representations of news content and user engagements. Yang et al.~\cite{yang2024update} proposed to extract domain-shared features from common users' engagements with news items to achieve detection performance improvement in both the data-sparser and the data-richer domains. However, the above mentioned cross-domain approaches fail to consider the impact of veracity-irrelevant features in news content on fake news detection and cross-domain knowledge transfer. In addition, most of them transfer knowledge only based on news content or user engagement. Though Mosallanezhad et al. utilized both of  news content and user engagement, they simply combines them without exploring engagement features of common users across domains for more effective knowledge transfer.

\section{Methodology}
\begin{figure*}[t]
  \centering
  \includegraphics[scale=0.222]{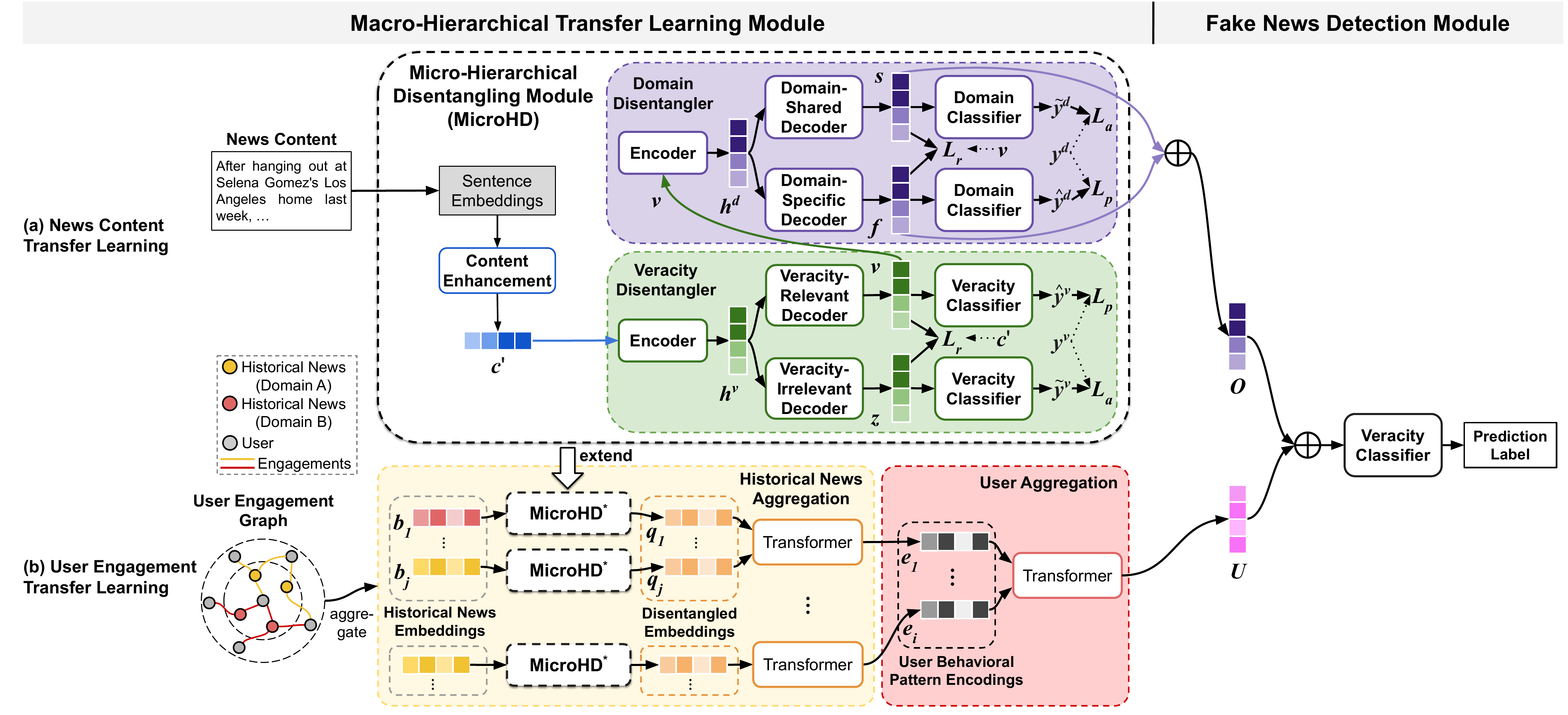}
  \caption{Overall framework of our MMHT}
  \label{fig:overall}
  \Description{overall structure}
\end{figure*}

\subsection{Problem Statement}
A fake news dataset records both news content and user engagements as well as the veracity label of each news item to indicate the true or fake. News content refers to main body of a news article, while user engagement refers to each user's set of engagements (e.g., reposting) with a historical news item in a certain time period. Specifically, dataset $\mathcal{C}_g=\{\mathcal{X}_1,\cdots, \mathcal{X}_n,\cdots,\mathcal{X}_{N}\}$ denotes a collection of news items within domain $g \in \mathcal{G}$, where $\mathcal{X}_n$ represents a news item, $N$ refers to the total number of news items in the collection, and $\mathcal{G} = \{\textit{political}, \textit{entertainment}\}$, where \textit{political} and \textit{entertainment} represent the political and entertainment news domains, respectively. Record of each news item $\mathcal{X}_n=\{\mathcal{O}_n, \mathcal{E}_n, y_n\}$ includes news article content $\mathcal{O}_n$, users' engagements with the news $\mathcal{E}_n$, and news veracity label $y_n \in \mathcal{Y}$. Specifically, $\mathcal{O}_n=\{o_1,\cdots,o_k\}$ represents a set of $k$ sentences in a news article, $\mathcal{E}_n=\{u_1,\cdots,u_i\}$ ($u_i \in \mathcal{U}$) represents $i$ users who have reposted news item $\mathcal{X}_n$, and $\mathcal{Y}=\mathcal{Y}_{true}\bigcup\mathcal{Y}_{fake}$, where $\mathcal{Y}_{true}$ denotes the set of true labels, and $\mathcal{Y}_{fake}$ denotes the set of fake labels. In addition, the record of a user $u_i$ includes a user-engaged historical news collection $\mathcal{B}_i=\{h_1,\cdots,h_j\}$ ($h_j\in\mathcal{H}$), which consists of $j$ pieces of historical news content that $u_i$ has engaged with. 

For each new item $\mathcal{X}_n$, given its article content $\mathcal{O}_n$ (i.e., \textit{news content aspect}), and $i$ engaged users as well as each user's $j$ engaged historical news items (i.e., \textit{user engagement aspect}), we aim to construct a cross-domain fake news detection framework $\mathcal{F}$ with three-fold goals: (1) disentangling veracity-relevant and domain-shared representations from $\mathcal{O}_n$ for effective knowledge transfer from news content aspect; (2) for each user in $\mathcal{E}_n$, including common and single-domain users, learning a representation of their behaviors from their corresponding $\mathcal{B}_i$, and further learning representation of target news' engagement pattern based on $\mathcal{E}_n$ for effective knowledge transfer from user engagement aspect, and (3) extracting veracity-relevant and domain-shared features based on both news content and user engagement from the source domain, and transferring them to the target domain to achieve better fake news detection performance in the target domain.

\subsection{Model Overview}
As shown in Fig. 3, our framework contains three modules: (1) \textit{micro-hierarchical disentangling module}, (2) \textit{macro-hierarchical transfer learning module}, and (3) \textit{fake news detection module}. Specifically, the module (2) is constructed on top of the module (1).

\subsection{Micro-Hierarchical Disentangling} \label{sec:micro}
We first present the structure of micro-hierarchical disentangling module in this subsection. 
Inspired by studies in~\cite{wang2024hierarchical}, we design this module to disentangle domain-shared and domain-specific features from veracity-relevant features, but this is different from studies in~\cite{wang2024hierarchical} in terms of both the disentangling objective and the neural network architecture.
As illustrated in Fig. 3, this module consists of two main components: (1) \textit{a content enhancement layer} to capture sentence-level relationships in news content utilizing co-attention mechanism~\cite{shu2019defend, li2020gcan}, (2) \textit{two serially connected disentanglers} to disentangle veracity-relevant and veracity-irrelevant features, as well as domain-shared and domain-specific features from news content, respectively. Both disentanglers are built on an adversarial auto-encoder framework~\cite{singhal2019spotfake}, which encompasses \textit{an encoder}, \textit{two decoders}, and a specifically designed \textit{loss function}. In this section, we delve into the specifics of the veracity disentangler, while the domain disentangler has a similar structure.

\subsubsection{Content Enhancement Layer} \label{sec:micro-content}
The content enhancement layer mainly contains a transformer network and a co-attention mechanism that takes the news content embedding as input. For a news item $\mathcal{X}_n$, we learn a content embedding $d_n$ based on the sentences in its article content:
\begin{equation}
    d_n=\operatorname{Dense}(d_n^o),
\end{equation}
where $\operatorname{Dense}$ indicates a fully connection network. $d_n^o$ is a concatenation of all sentence embeddings in news article content, which are initially generated using a high-performing pre-trained GPT-2 model~\cite{radford2019language}. Specifically, we perform fine-tuning to adapt the model to the requirement of our task. Each sentence embedding retains the default dimension of 768, corresponding to the output size of GPT-2's hidden states. Once content embedding $d_n$ is obtained, it is input into a transformer network to derive an embedding mask $mask_n$ for sentence embeddings in the content enhancement layer to generate two latent content embeddings $d_n^m$ and $d_n^b$:
\begin{equation}
    r_n^1=\operatorname{Transformer}(d_n),
\end{equation}
\begin{equation}
    r_n^2=\operatorname{LeakyReLU}(\operatorname{Dense}(r_n^1)),
\end{equation}
\begin{equation}
    mask_n=\operatorname{Sigmoid}(\operatorname{Dense}(r_n^2)),
\end{equation}
\begin{equation}
    d_n^m=d_n * mask_n, \; d_n^b=d_n * (1-mask_n),
\end{equation}

Upon obtaining the latent content embeddings $d_n^m$ and $d_n^b$, these two types of content embeddings serve as the input for the co-attention mechanism to generate context embedding $c_n$. To conserve the paper's length, we omit the detailed explanation of the co-attention mechanism. After context embedding $c_n$ is ready, this content embedding acts as the input for the veracity disentangler.

\subsubsection{Veracity-Irrelevant Decoder}
Right before the decoders in the veracity disentangler, there is \textit{an encoder} that takes the context embedding $c_n$ as input to derive the hidden representation $h_n^v$ for both veracity-relevant and veracity-irrelevant decoding:
\begin{equation}
    c_n^{\prime}= c_n + \operatorname{Dense}(d_n),
\end{equation}
\begin{equation}
    h_n^v= \operatorname{LeakyReLU}(\operatorname{Dense}(c_n^{\prime})).
\end{equation}
Besides, we employ residual connections to obtain the enhanced news content embeddings $c_n^{\prime}$.

For veracity-irrelevant decoder, it takes the hidden representation $h_n^v$ as the input and extracts the veracity-irrelevant representation $z_n$ from $h_n^v$ based on an attention network:
\begin{equation}
    p_n^1=\operatorname{softmax}\left(\frac{Q_n*K_n^\top}{\sqrt{d'}}\right) * V_n,
\end{equation}
\begin{equation}
    p_n^2=\operatorname{LeakyReLU}(\operatorname{Dense}(p_n^1)),
\end{equation}
\begin{equation}
    z_n=\operatorname{Sigmoid}(p_n^2) * h_n^v,
\end{equation}
where $\mathbf{Q_n}$ represents the query vector within the attention network, computed as $\mathbf{Q_n}=\mathbf{W_Q}*h_n^v+b_Q$. In this computation, $\mathbf{W_Q}$ is the weight matrix that projects the hidden representation $h_n^v$ into the query space, and $b_Q$ is the corresponding bias term that adjusts the projection. Similarly, $\mathbf{K_n}$ and $\mathbf{V_n}$ are derived from $h_v$ using transformations analogous to $\mathbf{Q_n}$, with their respective weight metrics and biases ($\mathbf{W_K}$, $b_K$ and $\mathbf{W_V}$, $b_V$). 
 
\subsubsection{Veracity-Relevant Decoder}
The veracity-relevant decoder consists of a decoder, which has an identical structure with the veracity-irrelevant decoder and is coupled with a news veracity classifier. This classifier takes the output of the decoder as input and predict the veracity label of the news item $\mathcal{X}_n$. During the training phase, we minimize the discrepancy between the predicted veracity label and the ground-truth label in the training dataset. By doing so, the veracity decoder is encouraged to extract the veracity-relevant feature effectively. Specifically, the veracity-relevant decoder takes the hidden representation $h_n^v$ as the input and generates the veracity-relevant representation $v_n$ using the same operations as described in Eqs. (8) to (10). Subsequently, the $v_n$ is fed to the veracity classifier to predict the veracity label $\hat{y}_n^v$:
\begin{equation}
    \hat{y}_n^v=\operatorname{softmax}(\operatorname{LeakyReLU}(\operatorname{Dense}(v_n))),
\end{equation}
where we employ the softmax function for classification. 

\subsubsection{Loss Function for Disentangling Module}
In order to effectively train the veracity disentangler, we designed three specific loss functions for its optimization: \textit{reconstruction loss}, \textit{veracity prediction loss}, and \textit{adversarial loss}. These three losses collectively contribute to the overall loss during the training phase. After the micro-hierarchical disentangling module is trained with test sets from both domains, its parameters are frozen and used for fake news detection.

The reconstruction loss plays a fundamental role in the training process of any auto-encoder structure. Therefore, we concatenate the disentangled veracity-irrelevant representation $z_n$ and the veracity-relevant representation $v_n$ to facilitate the reconstruction of the enhanced news content embeddings $c_n^{\prime}$. This ensures no information loss during the disentanglement process. Formally,
\begin{equation}
    L_r^v=([z_n;v_n]-c_n^{\prime})^2.
\end{equation}
The veracity prediction loss aims to guarantee that the disentangled veracity-relevant representation $v_n$ effectively captures the veracity information within the news content. Therefore, we employ the widely accepted binary cross-entropy loss $L_p^v$ to calculate this loss for each news item $\mathcal{X}_n$:
\begin{equation}
    L_p^v=-y_n^vlog(\hat{y}_n^v),
\end{equation}
where $y_n^v $ is the ground truth veracity label of $\mathcal{X}_n$. The adversarial loss is crucial for ensuring that the disentangled veracity-irrelevant representation $z_n$ remains entirely neutral with respect to any veracity information. To this end, we utilize $z_n$ to predict the veracity label $\tilde{y}_n^v$ of the target news $\mathcal{X}_n$ and subsequently aim to maximize the prediction error:
\begin{equation}
    L_a^v= \operatorname{max}(0,1+y_n^vlog(\tilde{y}_n^v)),
\end{equation}
where $\tilde{y}_n^v=\operatorname{softmax}(\operatorname{LeakyReLU}(\operatorname{Dense}(z_n))$.

Finally, overall loss of the veracity disentangler is the sum of all the aforementioned three losses:
\begin{equation}
    L_v=L_r^v+ L_p^v+L_a^v.
\end{equation}
Similarly, by adhering to the framework illustrated in Fig. 3, the overall loss of the domain disentangler can be calculated:
\begin{equation}
    L_d=L_r^d+ L_p^d+L_a^d.
\end{equation}

The total loss $L_m$ for micro-hierarchical disentangling module is calculated as:
\begin{equation}
    L_m=L_v+ \alpha_1 \cdot L_d,
\end{equation}
where $\alpha_1$ is the weight for domain disentangling loss.

\subsection{Macro-Hierarchical Transfer Learning}
The macro-hierarchical transfer learning module comprises two components: (1) the \textit{news content transfer learning} for knowledge transfer at news content aspect. In particular, this component is built upon the micro-hierarchical disentangling module and takes news content from both political and entertainment domains in the training phase to facilitate knowledge transfer across domains for generating news content representation; and (2) the \textit{user engagement transfer learning} to transfer domain-shared features of common users across domains at the user engagement aspect. This component utilizes user engagements as input to effectively model user behaviors and generate user engagement representation grounded in users' behaviors for the target news. Specifically, this component consists of two sequential layers: \textit{a historical news aggregation layer} and \textit{a user behavior aggregation layer}. As we have introduced the micro-hierarchical disentangling module in the previous section~\ref{sec:micro}, we here proceed to introduce the layers that compose the \textit{user engagement transfer learning} in this section.

\subsubsection{Historical News Aggregation}
As illustrated in Fig. 3, the historical news aggregation layer encompasses a Graph Convolutional Network (GCN) layer~\cite{zhang2023rumor, xiao2024msynfd}, a micro-hierarchical disentangling module trained in \textit{news content transfer learning} part, and a transformer network. This layer takes user engagements and content embeddings of historical news items as input. Specifically, given the user engagements $\mathcal{E}_n=\{u_1,\cdots,u_i\}$ of the target news item $\mathcal{X}_n$, we collect historical news collection $\mathcal{B}_i$ for each user $u_i$ in $\mathcal{E}_n$. Based on $\mathcal{E}_n$ and users' historical news collections $\{\mathcal{B}_1,\cdots,\mathcal{B}_i\}$, we construct a \textit{user engagement graph} $\mathcal{D}_n$ for $\mathcal{X}_n$ as depicted in Fig. 3, which includes two node types, i.e., news and users, and a single edge type presenting user engagements. In addition, we utilize the content embeddings of historical news items as the initial embeddings for news nodes in the graph. These embeddings are extracted in the same way as described in Section ~\ref{sec:micro-content}, while no predefined embeddings are assigned to user nodes. Once the graph $\mathcal{D}_n$ is obtained, we utilize a GCN layer to aggregate the content embedding $b_j$ for each historical news item in $\mathcal{B}_i$. Then, the historical news embedding sequence $\{b_1,\cdots,b_j\}$ is input into the historical news aggregation layer to learn behavioral pattern encodings $E_n$ for all users in $\mathcal{E}_n$:
\begin{equation}
    q_j=\operatorname{MicroHD^{\ast}}(b_j),
\end{equation}
\begin{equation}
    E_n=\operatorname{Transformer}(\{q_1,\cdots,q_j\}),
\end{equation}
where $MicroHD^{\ast}$ refers to the micro-hierarchical disentangling module trained in news content transfer learning. In order to extract users' behavioral patterns from their relevant historical news sequences, we employ a combination of GCN and Transformer. In our framework, we primarily rely on the transformer network to capture global information within the historical news sequences.

\subsubsection{User Behavior Aggregation}
To extract information from the behaviors of users who have engaged with the target news that can effectively determine the veracity of the news, in this layer, we learn user engagement representation for the target news from users' behavioral patterns. In particular, once we obtain the behavioral pattern encodings $E_n$, we utilize a GCN layer to aggregate the behavioral pattern encoding $e_i$ for each user who have engaged with the target news item $\mathcal{X}_n$. Next, the behavioral pattern encoding sequence $\{e_1,\cdots,e_i\}$ serves as the input for the user behavior aggregation layer to learn user engagement representation $U_n$ for the target news item $\mathcal{X}_n$:
\begin{equation}
    k_n^1=\operatorname{Transformer}(\{e_1,\cdots,e_i\}),
\end{equation}
\begin{equation}
    U_n=\operatorname{MLP}(k_n^1),
\end{equation}
where $MLP$ refers to a MLP layer which consists of a three-layer sequential structure: a linear layer, a LeakyReLU activation function, and another linear layer.

\subsection{Fake News Detection}
Both representation learned from news content and user engagement representation are integrated for improved fake news detection. First, the veracity-relevant domain-shared representation $s_n$ and veracity-relevant domain-specific representation $f_n$ are disentangled for each target news item $\mathcal{X}_n$, handling by the micro-hierarchical disentangling module in news content transfer learning. Subsequently, $s_n$ and $f_n$ are combined to build a complete news content representation $O_n$ with an MLP layer, as $O_n=MLP(s_n;f_n)$. At the same time, user engagement representation $U_n$ is learned from user engagements $\mathcal{E}_n$ and historical news collection $\mathcal{B}_n$ for each related user, handled by the user engagement transfer learning in macro-hierarchical transfer learning module. Finally, $O_n$ and $U_n$ are combined as the input to a veracity classifier for predicting the veracity label $y_n^{\prime}$ for each target news item $\mathcal{X}_n$.

The fake news detection loss $L$ is calculated based on the binary cross-entropy loss:
\begin{equation}
    L=-y_nlog(y_n^{\prime}),
\end{equation}
where $y_n$ is the ground truth veracity label of $\mathcal{X}_n$.

\section{Experiments and Analysis}
\subsection{Data Preparation}
We carried out evaluations on a commonly used news repository FakeNewsNet\footnote{https://github.com/KaiDMML/FakeNewsNet}~\cite{shu2020fakenewsnet}. The FakeNewsNet repository consists of two datasets: \textit{PolitiFact}, which contains news items from the political domain, and \textit{GossipCop}, which contains news items from the entertainment domain. In each dataset, for each news item, we utilize the news article as the news content, users' reposting records as user engagement, along with news domain labels and veracity labels.

Based on these datasets, we create a test set by randomly selecting $10\%$ of the news items, another $10\%$ for the validation set, and using the remainder for training. For each set, we separately construct a user engagement graph. In addition, we employ the widely adopted leave-one-out mechanism~\cite{han2022d, dong2020two} for training, testing and validation. Specifically, at each step, we select one news item as the target news and utilize the graph structure to identify its neighboring user nodes. Based on these user nodes, we identify other neighboring news nodes, excluding the target news, as the users' historical news items. Furthermore, we conduct 5-fold cross-validation and report the average results. The detailed characteristics of the experimental datasets are presented in Table~\ref{tab:characteristics}.

\begin{table}
  \setlength{\tabcolsep}{2mm}{
  \caption{The characteristics of two experimental datasets}
  \label{tab:characteristics}
  \small
  \begin{tabular}{lcc}
  \hline \hline {Statistics} & PolitiFact & GossipCop \\
  \hline 
  \# True News & 339 & 13,194 \\
  \# Fake News & 347 & 3,692 \\
  \hline 
  \# Single-Domain Users & 133,940 & 42,701 \\
  \# Common Users & 40,170 & 40,170 \\
  \hline 
  \# User Engagements & 419,358 & 225,467 \\
  \hline \hline
  \end{tabular}}
\end{table}

\begin{table*}
  \setlength{\tabcolsep}{3.4mm}{
  \caption{Comparison with baselines on two datasets (\%)}
  \label{tab:comparison_baselines}
  \small
  \begin{threeparttable}
  \begin{tabular}{l|cccc|cccc}
  \hline \hline \multirow{2}{*}{ Methods } & \multicolumn{4}{c|}{ PolitiFact } & \multicolumn{4}{c}{ GossipCop } \\
  \cline{2-9}
  & Prec. & Rec. & F1 & AUC & Prec. & Rec. & F1 & AUC \\
  \hline
dEFEND & $88.33 $ & $77.94 $ & $82.81 $ & $83.42 $ & $56.51 $ & $63.71 $ & $59.90 $ & $75.11 $ \\
BERT & $86.25 $ & $86.51 $ & $85.62 $ & $93.59 $ & $51.41 $ & $50.70 $ & $50.84 $ & $61.35 $ \\
GPT-4o mini & $67.59 $ & $66.11 $ & $63.84 $ & $66.11 $ & $57.32 $ & $58.28 $ & $43.13 $ & $58.28 $ \\
ARG & $88.00 $ & $88.47 $ & $88.06 $ & $92.57 $ & $75.08 $ & $73.81 $ & $74.08 $ & $86.39 $ \\
FANG & $61.19 $ & $62.02 $ & $61.83 $ & $79.18 $ & $51.58 $ & $52.12 $ & $50.50 $ & $62.59 $ \\
DUCK & $84.90 $ & $77.46 $ & $78.63 $ & $85.40 $ & $74.77 $ & $67.31 $ & $70.11 $ & $75.60 $ \\
\hline
REAL-FND & $53.25 $ & $63.59 $ & $57.96 $ & $68.19 $ & $56.70 $ & $71.17 $ & $63.12 $ & $68.50 $ \\
EANN & $54.09 $ & $63.52 $ & $46.73 $ & $62.17 $ & $64.67 $ & $66.06 $ & $64.65 $ & $66.25 $ \\
Amila et al. & $84.52 $ & $85.01 $ & $84.71 $ & $91.50 $ & $77.36 $ & $78.82 $ & $75.41 $ & $\underline{90.05} $ \\
M3FEND & $90.45 $ & $90.03 $ & $90.21 $ & $\underline{92.59} $ & $80.10 $ & $79.73 $ & $80.05 $ & $86.99 $ \\
UPDATE & $\underline{91.36}\tnote{2} $ & $\underline{90.56} $ & $\underline{90.91} $ & $92.16 $ & $\underline{81.16} $ & $\underline{80.05} $ & $\underline{80.92} $ & $87.21 $ \\
\hline
MicroHD & $90.44 $ & $90.86 $ & $90.56 $ & $94.97 $ & $82.39 $ & $75.19 $ & $77.83 $ & $87.54 $ \\
\textbf{MMHT} & $\mathbf{94.20}\tnote{1} $ & $\mathbf{94.52} $ & $\mathbf{94.33} $ & $\mathbf{97.40} $ & $\mathbf{87.09} $ & $\mathbf{82.68} $ & $\mathbf{83.53} $ & $\mathbf{92.18} $ \\
  \hline 
  \multicolumn{1}{c|}{\textit{Improvement}\tnote{3}} & $+3.1\%$ & $+4.4\% $ & $+3.8\% $ & $+5.2\% $ & $+7.3\%$ & $+3.3\% $ & $+3.2\% $ & $+2.4\% $\\
  \hline \hline
  \end{tabular}

  \begin{tablenotes}
  \footnotesize
  \item[1] A value in bold font indicates the best performance. 
  \item[2] An underlined value indicates the best performance of baseline approaches.
  \item[3] The improvement of MMHT over the best-performing baseline approaches, and the improvement is significant at $p<0.05$.
  \end{tablenotes}

  \end{threeparttable}
  }
\end{table*}

\subsection{Experiment Settings}
\subsubsection{Baseline Approaches}
The goal of our framework is to effectively transfer veracity-relevant and domain-shared features from the source domain to the target domain to improve its fake news detection performance, which leverages information from both news content and user engagement. To this end, we select two classes of news content-based and user engagement-based approaches as baselines: (1) single-domain fake news detection approaches, and (2) cross-domain fake news detection approaches. 
We select a total of 11 state-of-the-art and representative baseline approaches, with their details briefly outlined below.

\noindent \textbf{(1) Single-Domain Fake News Detection}:
\begin{itemize}
    \item \textbf{BERT}~\cite{devlin2018bert}: a transformer-based pre-trained language model that processes news content and is capable of being fine-tuned for fake news detection.
    \item \textbf{GPT-4o mini}: a lightweight large language model adapted from GPT-4 that processes news content and is capable of detecting fake news.
    \item \textbf{dEFEND}~\cite{shu2019defend}: a model that jointly attends to interactions between news content and user comment to predict news veracity label.
    \item \textbf{ARG}~\cite{hu2024bad}: a model that integrates knowledge from a large language model into a fine-tuned small language model to process news content.
    \item \textbf{FANG}~\cite{nguyen2020fang}: a model that learns graph representations from social context information such as user engagement.
    \item \textbf{DUCK}~\cite{tian2022duck}: a model that leverages three graph-based features derived from user engagement.
\end{itemize}
\textbf{(2) Cross-Domain Fake News Detection}:
\begin{itemize}
    \item \textbf{EANN}~\cite{wang2018eann}: an adversarial learning-based method that extracts domain-shared features from information such as news content.
    \item \textbf{Amila et al.}~\cite{silva2021embracing}: an adversarial learning-based model that extracts both domain-shared and domain-specific features from information such as news content.
    \item \textbf{M3FEND}~\cite{zhu2022memory}: a model that leverages three types of textual features from news content and enriches their domain information by modeling domain discrepancies.
    \item \textbf{Real-FND}~\cite{mosallanezhad2022domain}: a reinforcement learning-based model that extracts domain-shared features from news content and user engagement.
    \item \textbf{UPDATE}~\cite{yang2024update}: a model that achieves cross-domain knowledge transfer based on user engagement and integrates news content for cross-domain fake news detection.
\end{itemize}
\subsubsection{Evaluation Metrics}
To evaluate the effectiveness of our framework in cross-domain fake news detection, we compare the performance of the proposed framework MMHT not only with state-of-the-art single-domain methods but also with state-of-the-art cross-domain approaches. In addition, to evaluate the performance of framework in disentangling news content representation, we solely use the combination of representations $s_n$ and $f_n$ extracted by the micro-hierarchical disentangling module (MicroHD) for fake news detection and compare the results with those of state-of-the-art single-domain content-based fake news detection approaches. 
In order to provide a comprehensive evaluation, we utilize four commonly used metrics: Precision (Prec), Recall (Rec), F1-score (F1), and AUC. In addition, following~\cite{wang2022veracity}, we conduct a significance test for all the metrics using a paired t-test.
\subsubsection{Parameter Settings}
For fair comparison, we carefully optimize all model parameters and hyperparameters for both the baseline approaches and our proposed framework using the validation set. For each baseline approach, we initialize parameters according to the values reported in the original paper and then fine-tune them on our datasets to achieve optimal performance. In our model, we set the veracity loss weight and domain loss weight to 1 and 0.1, respectively, for both \textit{PolitiFact} and \textit{GossipCop} datasets. We employ separate transformers for historical news aggregation and user behavior aggregation, each configured with 2 attention heads. Additionally, we utilize 20 records of user engagements for each target news, and for each user, we consider 20 historical news items to generate user features. We train for 20 epochs on the \textit{PolitiFact} dataset and 10 epochs on the \textit{GossipCop} dataset. Furthermore, we assess the sensitivity of MMHT on three critical parameters, which is presented in Section A in the appendix.

\begin{table}
  \setlength{\tabcolsep}{0.3mm}{
  \caption{Comparison with the variants on two datasets (\%)}
  \label{tab:variants}
  \small
  \begin{threeparttable}
  \begin{tabular}{l|cccc|cccc}
  \hline \hline \multirow{2}{*}{ Methods } & \multicolumn{4}{c|}{ PolitiFact } & \multicolumn{4}{c}{ GossipCop } \\
  \cline{2-9}
  & Prec & Rec & F1 & AUC & Prec & Rec & F1 & AUC \\
  \hline 
    MMHT & $\mathbf{94.20} $ & $\mathbf{94.52} $ & $\mathbf{94.33} $ & $\mathbf{97.40} $ & $\mathbf{87.09} $ & $\mathbf{82.68} $ & $\mathbf{83.53} $ & $\mathbf{92.18} $ \\
    MMHT-SG & $91.73 $ & $91.84 $ & $91.78 $ & $94.73 $ & $82.30 $ & $80.01 $ & $80.47 $ & $87.01 $ \\
    MMHT-M & $93.00 $ & $93.11 $ & $93.05 $ & $96.51 $ & $86.69 $ & $80.15 $ & $80.73 $ & $92.02 $ \\
    \hline
    MicroHD & $\mathbf{90.44} $ & $\mathbf{90.86} $ & $\mathbf{90.56} $ & $\mathbf{94.97} $ & $\mathbf{82.39} $ & $\mathbf{75.19} $ & $\mathbf{77.83} $ & $\mathbf{87.54} $ \\
    MicroHD-V & $88.13 $ & $87.76 $ & $87.92 $ & $93.89 $ & $77.40 $ & $73.98 $ & $75.97 $ & $81.63 $ \\
    MicroHD-D & $89.07 $ & $88.18 $ & $88.49 $ & $95.11 $ & $77.17 $ & $74.01 $ & $76.09 $ & $80.57 $ \\
  \hline \hline
  \end{tabular}
  \end{threeparttable}
  }
\end{table}

\subsection{Performance Comparison with Baselines}
The main results are shown in Table~\ref{tab:comparison_baselines}. We can clearly find that our proposed MMHT significantly outperforms the best-performing baseline approaches over all metrics across two datasets. Specifically, our MMHT improves the performance by an average of $4.13\%$ across all metrics, with improvements up to $5.2\%$ in AUC on \textit{PolitiFact}. On \textit{GossipCop}, our MMHT achieves an average performance increase of $4.05\%$ across all metrics, with a maximum improvement of $7.3\%$ in precision. Furthermore, on the one hand, our MMHT outperforms single-domain baseline approaches on both datasets. This result demonstrates that, by leveraging cross-domain knowledge transfer, our MMHT achieves the best performance in fake news detection on both datasets.
On the other hand, our MMHT surpasses cross-domain baseline approaches. This result indicates that, by utilizing both news content and user engagement for knowledge transfer, our MMHT achieves greater improvement compared to other cross-domain approaches on both datasets.

In addition, we evaluate the performance of MicroHD module in MMHT by comparing its results for fake news detection with those of single-domain news content-based baseline approaches. The results are shown in Table~\ref{tab:comparison_baselines}. It is evident that the performance of our MicroHD in fake news detection outperforms that of single-domain news content-based baseline approaches on both datasets. 
This result demonstrates that, by disentangling veracity-relevant and veracity-irrelevant features before extracting domain-shared and domain-specific features, our MicroHD effectively mitigates the negative impact of veracity-irrelevant feature for fake news detection and achieves better detection performance than the content-based baseline approaches.
In particular, our MicroHD improves the performance by an average of $2.35\%$ across all metrics on \textit{PolitiFact}, with improvements up to $2.8\%$ in F1. On \textit{GossipCop}, it achieves an average improvement of $4.50\%$ across all metrics, with improvements up to $9.7\%$ in precision.

\subsection{Ablation Study}
\subsubsection{Settings}
To evaluate the rationale and effectiveness of the designed modules in our framework, we conduct an ablation study, where we compare the performance of MMHT with its two variants:
(1) \textbf{MMHT-SG}: which removes the cross-domain mechanism within both news content-aspect and user engagement-aspect transfer learning. Specifically, it trains MicroHD using news samples from a single domain in news content aspect and removes common user in user engagement aspect (i.e., a single-domain setting);
(2) \textbf{MMHT-M}: which removes the trained MicroHD in user engagement transfer learning by directly utilizing initial historical news embedding sequence $\{b_1,\cdots,b_j\}$ as the input of transformer in historical news aggregation.
In addition, we further compare the performance of MicroHD within MMHT with its two variants. To isolate the performance improvements brought by MicroHD, we specifically use the variants of MicroHD for this comparison.
(1) \textbf{MicroHD-V}: which removes the veracity disentangler in MicroHD by directly taking the enhanced news content embeddings $c_n^{\prime}$ as the input of domain disentangler;
(2) \textbf{MicroHD-D}: which removes the domain disentangler in MicroHD by directly replacing its outputs $s_n$ and $f_n$ with the output of veracity disentangler $v_n$.

\subsubsection{Performance Comparison with Variants}
The results in Table~\ref{tab:variants} demonstrate the following findings:
(1) \textit{Our MMHT effectively improves fake news detection performance in both data-sparser and data-richer domain}. We observe a numerical decline across all metrics when training our MMHT under the single-domain setting (i.e., \textbf{MMHT-SG}). 
This result highlights that, in our MMHT, the cross-domain knowledge transfer based on news content and user engagement is effective, and the detection performance in both domains is notably improved.
%the effectiveness of our MMHT in cross-domain fake news detection.
(2) \textit{Our MicroHD can improve the performance of cross-domain fake news detection}. As shown in Table~\ref{tab:variants}, removing the trained MicroHD from user engagement transfer learning (i.e., \textbf{MMHT-M}) leads to a notable decline in fake news detection performance on both datasets.
(3) \textit{Both the veracity and domain disentanglers enhance the accuracy of news content representation}. By comparing the performance of \textbf{MicroHD} with its two variants (i.e., \textbf{MicroHD-V} and \textbf{MicroHD-D}), we find that removing either the veracity disentangler or the domain disentangler results in a decline in the accuracy of news content representation. This result indicates that our MicroHD is able to effectively disentangle veracity-relevant and veracity-irrelevant features, as well as domain-shared and domain-specific features. 

\subsection{Case Study}
To provide an illustrative demonstration
of feature categories in news content and user engagement, we conducted a case study on \textit{PolitiFact} and \textit{GossipCop} datasets, which is presented in Section B in the appendix.

\section{Conclusions and Future Work}
In this paper, we propose a novel macro- and micro-hierarchical transfer learning framework for cross-domain fake news detection. From the micro perspective, we propose a novel micro-hierarchical disentangling module to address the negative impact of veracity-irrelevant features in news content for cross-domain fake news detection. From the macro perspective, we propose a novel macro-hierarchical transfer learning module to further disentangle veracity-irrelevant features in user engagement for more accurate user engagement features in the source domain, thereby enabling improved detection performance in the target domain. Extensive experiments on real-world datasets demonstrate the superiority of our framework over the state-of-the-art single-domain and cross-domain fake news detection approaches. Future work will focus on exploring more effective approaches to disentangle news veracity and domain information from news content, aiming to further improve the performance of cross-domain fake news detection.

\begin{acks}
This research is supported by ARC Discovery Projects DP200101441 and DP230100676, Australia, the National Research Foundation, Singapore and Infocomm Media Development Authority under its Trust Tech Funding Initiative. Any opinions, findings and conclusions or recommendations expressed in this material are those of the author(s) and do not reflect the views of National Research Foundation, Singapore and Infocomm Media Development Authority
\end{acks}

% \bibliographystyle{ACM-Reference-Format}
\balance
% \bibliography{mybase}


\clearpage
\appendix

\begin{figure}[tbp]
    \centering
    \begin{subfigure}[b]{0.22\textwidth}
        \centering
        \includegraphics[width=\textwidth]{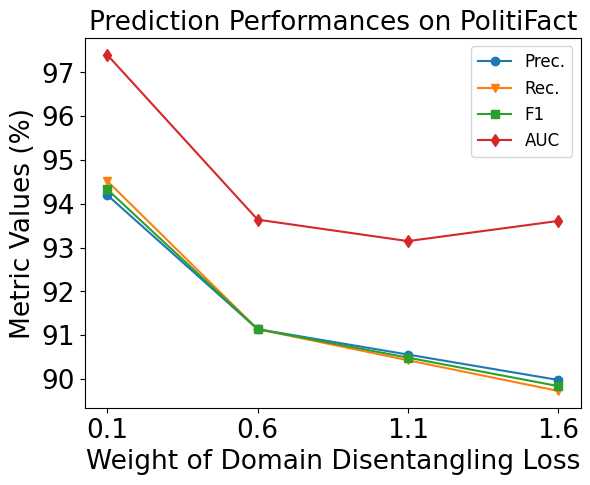}
        \label{fig:dloss_p}
    \end{subfigure}
    \hfill
    \begin{subfigure}[b]{0.22\textwidth}
        \centering
        \includegraphics[width=\textwidth]{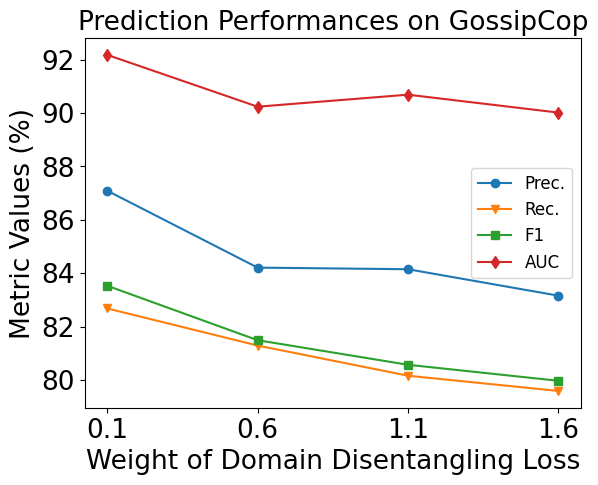}
        \label{fig:dloss_g}
    \end{subfigure}
    \caption{The impact of weight of domain disentangling loss}
    \label{fig:dloss}
\end{figure}

\begin{figure}[tbp]
    \centering
    \begin{subfigure}[b]{0.22\textwidth}
        \centering
        \includegraphics[width=\textwidth]{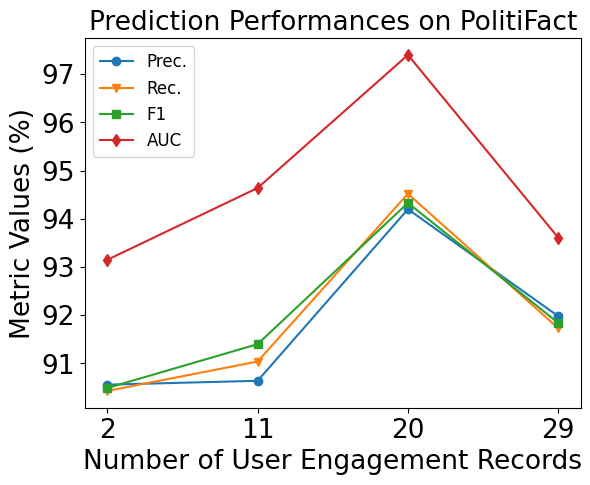}
        \label{fig:engagement_p}
    \end{subfigure}
    \hfill
    \begin{subfigure}[b]{0.22\textwidth}
        \centering
        \includegraphics[width=\textwidth]{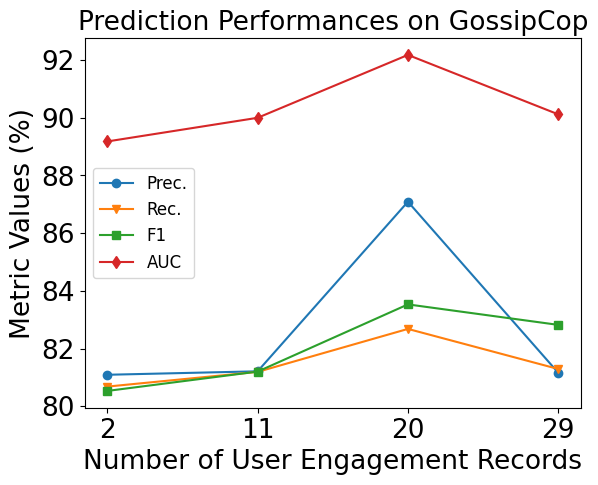}
        \label{fig:engagement_g}
    \end{subfigure}
    \caption{The impact of number of user engagement records}
    \label{fig:engagement}
\end{figure}

\begin{figure}[tbp]
    \centering
    \begin{subfigure}[b]{0.22\textwidth}
        \centering
        \includegraphics[width=\textwidth]{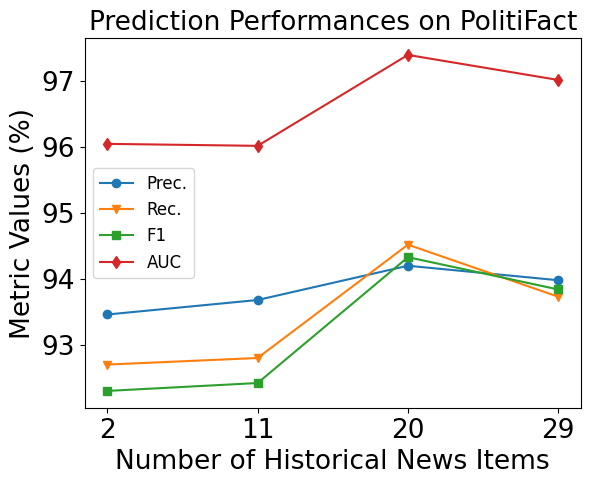}
        \label{fig:historical_p}
    \end{subfigure}
    \hfill
    \begin{subfigure}[b]{0.22\textwidth}
        \centering
        \includegraphics[width=\textwidth]{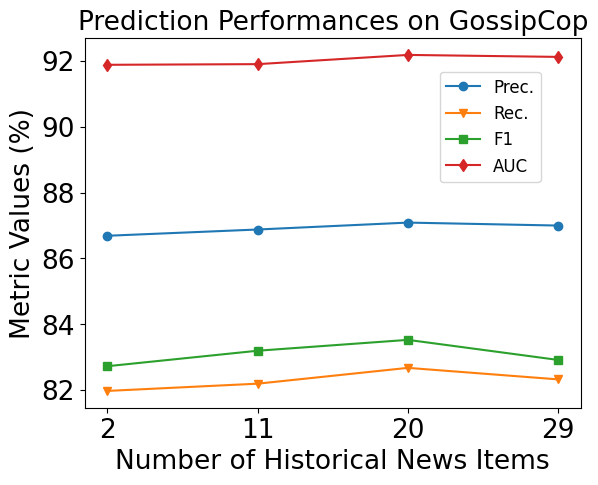}
        \label{fig:historical_g}
    \end{subfigure}
    \caption{The impact of number of historical news items}
    \label{fig:historical}
\end{figure}

\section{Parameter Analysis}
\subsection{Performance w.r.t. the weight of domain loss}
We examine the influence of the domain disentangling loss weight $\alpha_1$ on detection performance across both \textit{PolitiFact} and \textit{GossipCop} datasets. The value of $\alpha_1$ is varied from 0.1 to 1.1 with an increment step of 0.5. 

As depicted in Fig.~\ref{fig:dloss}, the optimal performance is observed when $\alpha_1$ is set to 0.1. Increasing values of $\alpha_1$ negatively impacts fake news detection performance, as it overemphasizes domain features while neglecting the extraction of veracity-related features.

\subsection{Performance w.r.t. the number of user engagement records and historical news items}
We investigated the influence of varying the number of user engagement records and historical news items on model performance across both datasets. The values were systematically adjusted from 2 to 29 with a step size of 9. To analyze the impact, we fixed the value of one variable while varying the other, allowing us to compare how changes in each factor affect performance. As shown in Fig.~\ref{fig:engagement} and Fig.~\ref{fig:historical}, the optimal configuration is achieved when the number of user engagement records is set to 20 and historical news items is set to 20. Fewer user engagement records or historical items fail to provide sufficient contextual information, while higher numbers introduce redundant data, increasing computational overhead and negatively impacting the detection performance.

\section{Case Study}
\begin{figure*}[t]
  \centering
  \includegraphics[scale=0.25]{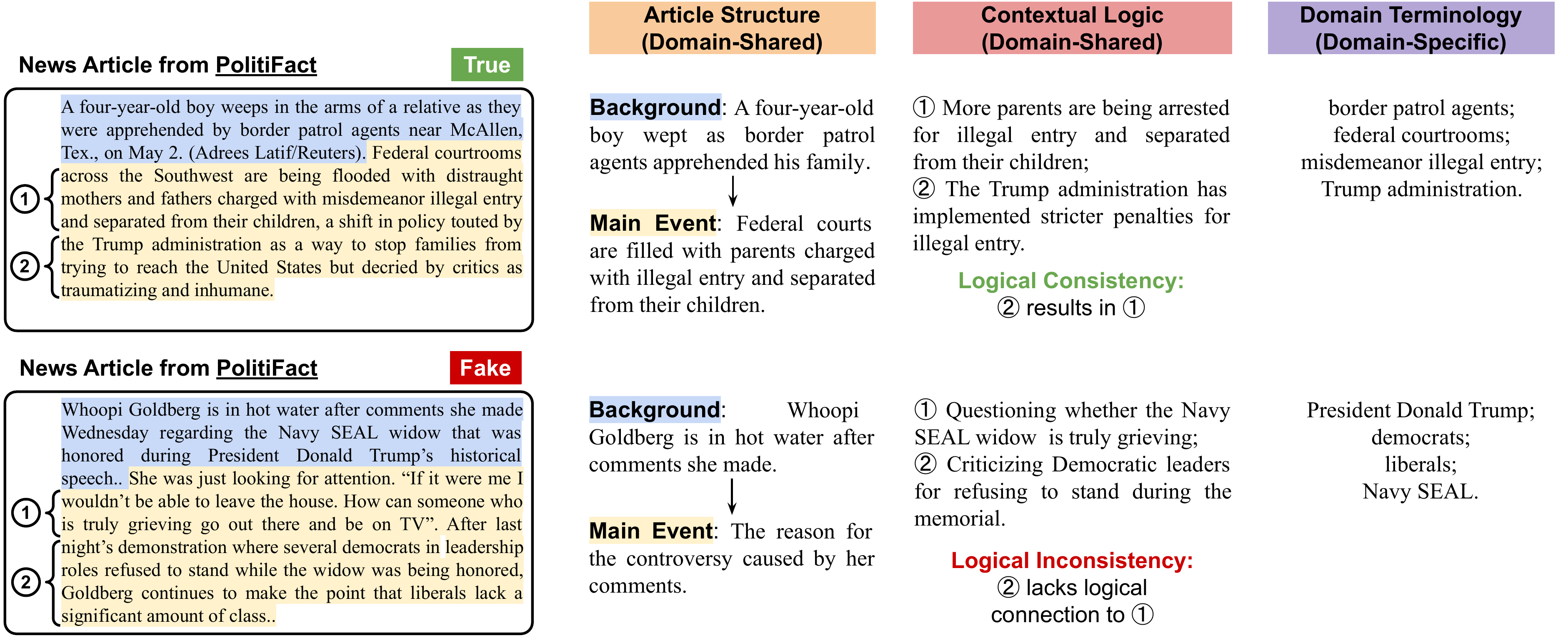}
  \caption{Case studies of news article from PolitiFact.} 
  \label{fig:case_study-1}
  \Description{case study of PolitiFact}
\end{figure*}

\begin{figure*}[t]
  \centering
  \includegraphics[scale=0.25]{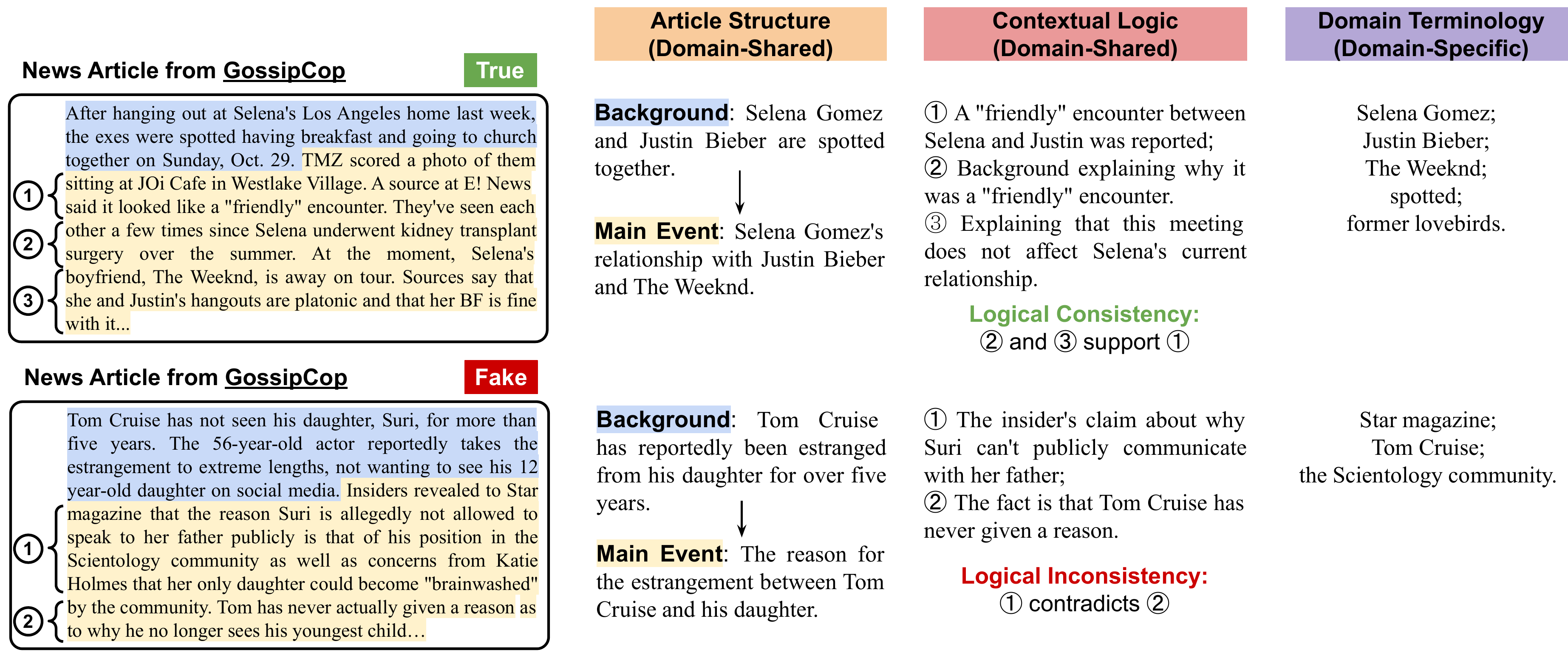}
  \caption{Case studies of news article from GossipCop.} 
  \label{fig:case_study-2}
  \Description{case study of GossipCop}
\end{figure*}

We present four representative examples from both PolitiFact and GossipCop datasets that are engaged by a common user in Fig.~\ref{fig:case_study-1} and Fig.~\ref{fig:case_study-2}. We also selected three features of news content as examples: article structure, contextual logic, and domain terminology. 

From the case studies, when comparing news from different domains, article structure and contextual logic are considered domain-shared features, as different domains exhibit similarities in these features. In contrast, domain terminology is a domain-specific feature because the categories of terminology vary across domains. Furthermore, when comparing true and fake news, article structure is a veracity-irrelevant feature since fake news often mimics the structure of true news. In contrast, contextual logic is a veracity-relevant feature and thus it can help effectively distinguish between true and fake news.

\end{document}